\documentclass[a4paper,12pt]{article}
\usepackage{amsmath,amsthm,amssymb,epsf,epsfig}
\usepackage{array}
\usepackage{cite}

\newcommand{\remark}[1]{}
\let\LARGE=\Large
\let\Large=\large

\makeatletter
\@addtoreset{equation}{section}
\makeatother

\makeatletter
\@addtoreset{equation}{section}
\makeatother

\begin{document}

\begin{flushright}
QMUL-PH-04-11\\
\end{flushright}

\vspace{8truemm}
\begin{centerline}
{\bf \LARGE D-Brane Dynamics and NS5 Rings }
\end{centerline}
\bigskip
\vspace{4truecm}
\centerline{ \bf Steven Thomas\footnote{e-mail:
 s.thomas@qmul.ac.uk} \
and John ward\footnote{e-mail:
 j.ward@qmul.ac.uk} }
\medskip

\begin{center}
{\em Department of Physics,\\
Queen Mary, University of London,\\
Mile End Road,\\
London E1 4NS, U.K.\\}

\end{center}

\vspace{2truecm}


\begin{abstract}
We consider the classical motion of a probe D-brane moving in the background geometry of a 
ring of NS5 branes, assuming that the latter are non-dynamical. We analyse the solutions to the 
Dirac-Born-Infield (DBI) action governing the approximate dynamics of the system.  
In the near horizon (throat) approximation we find several exact solutions for the probe brane motion. 
These are compared to numerical solutions obtained in more general cases. One solution of particular 
interest is when the probe undergoes oscillatory motion through the centre of the ring (and perpendicular to it).
By taking the ring radius sufficiently large, this solution should remain stable to any stringy
corrections coming from open-strings stretching between the probe and the NS5-branes along the ring.

\end{abstract}


\newpage

\section{Introduction.}
There has been a flourish of recent work concerned with time dependence in
string theory ( see, for example \cite{kutasov},\cite{pqstring}, \cite{sahayakan}, \cite{Ghodsi} and 
references therein), where we have a $Dp$-brane probing the backgrounds of various supergravity solutions.
Whilst some of this work has focused on $D$-brane backgrounds (\cite{burgess}, \cite{Panigrahi} ), much
of it has dealt with $NS$5-branes. This has been particularly fruitful since it is known that the
instability of the probe brane can be explained using tachyon rolling \cite{sen}, and has also emphasised
the relationship betweeen BPS and non-BPS branes \cite{kutasov2}, \cite{kluson}.

In a recent paper \cite{kutasov}, the dynamics of a $D$-brane in the background of
$k$ coincident $NS5$-branes was considered using the DBI action.
The $D$-brane is effectively a probe, since it can explore the background without disturbing it.
 This is because the fivebrane tension
goes as $1/g_s^2$, whilst the $D$-brane tension only goes as $1/g_s$. Thus the
$NS$5-branes are much heavier in the $g_s\to 0$ limit. Geometrically this
implies that the fivebranes form an infinite throat in spacetime
where the string coupling increases as we move closer to the bottom.

Since the $D$-brane and the $NS5$-brane
preserve different halves of the type II supersymmetry, this configuration
is unstable despite the fact that the $D$-brane is BPS. This means that
the probe brane will be gravitationally
attracted towards the sources. The Ramond-Ramond charge of the $Dp$-brane
is free to leak to the fivebranes during the motion, and we also find that
the energy due to string emission is divergent for $p \le 2$ \cite{sahayakan}. This
indicates that the classical analysis is vaild only for larger dimension
branes. As the brane loses almost all of its energy it can form a bound
state with the fivebranes, becoming a $(k,1)$ fivebrane - a member of the
$SL(2,\mathbf{Z})$ fivebrane multiplet.

The issue of stable trajectories was also discussed in \cite{kutasov},
but it was found that the probe brane was either scattered or doomed to
fall towards the $NS$5-branes. However this was of interest since
the infalling $D$-brane would probe the strong coupling region, where
pertubation theory would be expected to fail. It was argued that there
was a range of energies for which pertubation theory was still reliable
as the probe passed through the thoat.

In this note we would like to examine a similar situation, however rather
than considering $k$ coincident $NS$5-branes we will be interested in a ring
of $NS$5-branes. As in \cite{kutasov} this also has an exact
description in terms of a coset model CFT (\cite{sfetsos},\cite{sfetsos2})
 $SU(2)/U(1) \times SL(2,\mathbf{R})/U(1) $.
This background solution will also provide us with a
throat geometry to probe, which is more complicated than in the point-like case.
In particular it is possible to find solutions that explore this throat region but without the string coupling 
blowing up and  as such we can expect that perturbative techniques will
be valid in this case. In addition we also couple
an electric field to the $D$-brane worldvolume and investigate the interplay
between electrical energy and the energy density, and consider how this
affects the probe brane dynamics. Since our configuration will appear
pointlike at large distances, we expect some overlap with the work done in \cite{kutasov}.
We are interested in whether this picture will lead to the formation of bound orbits, which is of particluar interest
for cosmology (see \cite{mirage}, \cite{tak} and references therein), and also the possible 
implications for tachyon rolling.

We begin the note with a review of the supergravity background of the
problem and introduce the effective $D$-brane action. We then go on to
discuss the dynamics of the probe brane in the different limits of the
theory before concluding with some remarks.

\section{$NS5$-brane background and D-brane action.}
We wish to study the dynamics of a $Dp$-brane in the background gravitational 
potential generated by a ring of $k$, static $NS5$ branes in type II string theory. In order to do this
we must first consider the background fields which are given by the CHS solution \cite{CHS}.
We find that the metric, dilaton and NS $B$-field are

\begin{eqnarray} \label{eq:CHS}
ds^2&=& dx_{\mu} dx^{\mu} + H(x^n) dx^m dx^m\nonumber\\
\frac{g_s^2 (\phi)}{g_s^2} = e^{2(\phi-\phi_0)}&=&H(x^n)\nonumber \\
H_{mnp}&=&-\epsilon^q_{mnp} \partial_q \phi
\end{eqnarray}

where  $\mu = 0 ..5$ label the coordinates parallel to the NS5-brane  and the indices $m$ run 
over the four transverse dimensions
and $g_s $ is the string coupling in the asymtotically flat region.
As usual $H_{mnp}$ is the $3$-form NS field strength, and $H(x^n)$ is the harmonic function describing the $NS5$ 
branes which satisfies the Poisson equation in the transverse space. For fivebranes at arbitrary positions 
with respect to an origin, the harmonic function is descibed by
\begin{equation} 
H = 1 + l_s^2 \sum_{i=1}^k  \frac{1}{\vert \bf{x}- \bf{x}_i \vert ^2}
\end{equation}
where $l_s = \sqrt{\alpha'}$ is the string length. The geometry under consideration
in this note is obtained from the extremal limit of the rotating $NS5$ brane solutions discussed in 
\cite{sfetsos}, with the harmonic function  given by

\begin{equation}\label{eq:harmonic1}
H= 1 + \frac{k l_s^2}{\sqrt{(l_1^2-l_2^2+x_6^2+x_7^2+x_8^2+x_9^2)^2-4(l_1^2-l_2^2)(x_6^2+x_7^2)}}.
\end{equation}
This represents a continuous uniform distribution of branes along a ring of radius
$R=\sqrt{|l_1^2-l_2^2|}$, which is oriented in the $x_6-x_7$ plane in the transverse
Euclidean space. It will be convenent to make the following coordinate transformations
to fully exploit the symmetry of the problem:
\begin{eqnarray}\label{eq:coords}
x_6 & = & \rho \, {\rm cos}(\theta), \hspace{0.5cm} x_7 = \rho \, {\rm sin}(\theta) \\
x_8 & = & \sigma \,{\rm cos}(\phi), \hspace{0.5cm} x_9 = \sigma \,{\rm sin}(\phi)
\end{eqnarray}
which effectively defines an $SO(2) \times SO(2)$ symmetry.
In terms of these coordinate definitions, the harmonic function reduces to
\begin{equation}\label{eq:harmonic2}
H(\rho, \sigma) = 1 + \frac{k l_s^2}{\sqrt{(R^2+\rho^2 + \sigma^2)^2 - 4R^2\rho^2}}.
\end{equation}
We now wish to introduce our probe $Dp$-brane into this background which is described
by the effective DBI action. We assume that the worldvolume of the probe fills out
the directions $x_1 \ldots x_p$, and use the reparameterization invariance present to
go to static gauge. The transverse directions to the $NS5$ branes induce scalar
fields on the $Dp$-brane world volume, whose behaviour is described by the DBI action
\begin{equation}\label{eq:DBI}
S = -\tau_p \int d^{p+1} \zeta e^{(\phi-\phi_0)} \sqrt{-det(G_{\mu \nu}+B_{\mu \nu}+
2\pi l_s^2 F_{\mu \nu})}.
\end{equation}
$\tau_p$ is the $Dp$-brane tension, $F_{\mu\nu}$ is the $U(1)$ gauge field, whilst $G_{\mu \nu}$
and $B_{\mu \nu}$ are the pullbacks of the metric and the $B$ field to the brane:
\begin{equation}\label{eq:metric}
G_{\mu \nu}= \partial_{\mu} X^A \partial_{\nu}^B G_{AB}(X)
\end{equation}
\begin{equation}\label{eq:Bfield}
B_{\mu\nu}= \partial_{\mu} X^A \partial_{\nu}^B B_{AB}(X)
\end{equation}
where $A, B = 0, 1, 2, \ldots , 9$ run over the ten dimensional bulk spacetime and
$G_{AB}, B_{AB}$ are the bulk metric and $B$-field. 

\section{Dynamics of the probe brane.}

For the rest of this note we will be interested in homogenous solutions of the 
equations of motion, where the transverse scalars are time dependent only. This
will also ensure that coupling to the B field vanishes and so we will ignore it from now on.
We will also consider a non zero electric field (only) on the brane world volume, i.e
$F_{0m}=E_m$. So, the induced metric becomes
\begin{equation}\label{eq:metric2}
G_{\mu\nu} = \eta_{\mu\nu} + H(x^n) \delta_{\mu}^0 \delta_{\nu}^0 
 (\dot{\rho}^2+ \rho^2 \dot{\theta}^2 + \dot{\sigma}^2 + \sigma \dot{\phi}^2)
\end{equation}
and upon substitution into (\ref{eq:DBI}) we obtain
\begin{equation} \label{eq:action}
S=-\tau_p \int d^{p+1} \zeta \sqrt{H^{-1}-\dot{\rho}^2-\dot{\sigma}^2-\rho\dot{\theta}^2-
\sigma\dot{\phi}^2-H^{-1} F^2}
\end{equation}
where we have defined  $F^2= E^m E_m$ as the constant electric field strength in dimensionless 
units by absorbing the factors of $l_s$. 
Also we have assumed (see \cite{tak} ) that $F^2 $ is small in order to obtain the relatively
simple form of the action (\ref{eq:action}).

From the action we can also deduce the following canonical momenta:
\begin{equation}
\Pi_\sigma = \frac{m\dot{\sigma}}{\sqrt{H^{-1}(1-F^2)-(\dot{\rho}^2+\dot{\sigma}^2
+\rho^2 \dot{\theta}^2 + \sigma^2 \dot{\phi}^2)}}
\end{equation}
\begin{equation}
\Pi_\rho = \frac{m\dot{\rho}}{\sqrt{H^{-1}(1-F^2)-(\dot{\rho}^2+\dot{\sigma}^2
+\rho^2 \dot{\theta}^2 + \sigma^2 \dot{\phi}^2)}}
\end{equation}
\begin{equation}
L_{\theta}= \frac{m\rho^2 \dot{\theta}}{\sqrt{H^{-1}(1-F^2)-(\dot{\rho}^2+\dot{\sigma}^2
+\rho^2 \dot{\theta}^2 + \sigma^2 \dot{\phi}^2)}}
\end{equation}
\begin{equation}
L_{\phi} = \frac{m \sigma^2\dot{\phi}}{\sqrt{H^{-1}(1-F^2)-(\dot{\rho}^2+\dot{\sigma}^2
+\rho^2 \dot{\theta}^2 + \sigma^2 \dot{\phi}^2)}}
\end{equation}
where $m= \tau_p \int d^p \zeta$ represents the effective `mass' of the brane.
It will be useful to rescale these momenta to remove this mass dependence,
and so we are left with:
\begin{equation}
{\tilde{\Pi}}_\rho =\Pi_\rho /m,\hspace{0,2cm} {\tilde{\Pi}}_\sigma= \Pi_\sigma /m,
\end{equation}
\begin{displaymath}
\tilde{L}_{\theta}=L_{\theta}/m,\hspace{0.2cm} \tilde{L}_{\phi}=L_{\phi}/m.
\end{displaymath}

It now becomes a fairly straightforward procedure to calculate the canonical
energy density of the brane.
\begin{equation}\label{eq:energy}
\tilde{E} \equiv \frac{E}{m} =
 \frac{1}{H\sqrt{(H^{-1}(1-F^2)-(\dot{\rho}^2+\dot{\sigma}^2+ \rho^2 \dot{\theta}^2 + \sigma^2 \dot{\phi}^2)}},
\end{equation}
from which we obtain the following equation for the  motion of $\rho$ and $\sigma$:
\begin{equation}\label{eq:eom}
(\dot{\rho}^2+\dot{\sigma}^2) = \frac{(1-F^2)}{H(\rho,\sigma )}-\frac{1}{H^2(\rho,\sigma )\tilde E^2} 
\left( 1+\frac{\tilde{L}_{\theta}^2}{\rho^2}+\frac{\tilde{L}_{\phi}^2}{\sigma^2} \right)
\end{equation}
Since the RHS of this equation is non-negative it imposes constraints on the strength of the electric field. We first
set the angular momentum terms to zero, which means that the following constraint must be satisfied:
\begin{equation} \label{eq:constraint1}
H(1-F^2) \ge 1/\tilde{E}^2.
\end{equation}
In order to determine the full constraint it is necessary to specify the trajectory of the probe brane. Since there
is an explicit $SO(2)\times SO(2)$ symmetry we can choose to investigate the motion in the plane parallel to the
ring (i.e $\sigma=0$), or in the plane transverse to the ring (i.e $\rho=0$). In the first instance, setting
$\sigma$ to zero reduces the harmonic function (\ref{eq:harmonic2}) to
\begin{equation}\label{eq:planeharmonic}
H(\rho)=1+\frac{kl_s^2}{|R^2-\rho^2|}.
\end{equation}
Which can easily be seen to be singular at $\rho=R$ when the probe brane hits the ring. Upon insertion into the contraint
equation we find:
\begin{equation}
\frac{kl_s^2}{|R^2-\rho^2|}(1-F^2)-F^2  \ge \frac{1}{\tilde{E}^2}-1.
\end{equation}
There are three cases to consider. Firstly we have $1=\tilde{E}$, which means that the RHS will vanish identically
and so $F^2$ must be less than unity. In the limit of small $F^2$ which we are working this is automatically satisfied.
 These conditions are also the same in the $1 > \tilde{E}$ case and in the $1 < \tilde{E}$ cases.
 
We can also consider the case of motion transverse to the disk plane by fixing  $\rho=0$. This gives us a new
harmonic function, namely
\begin{equation} \label{eq:perpharmonic}
H(\sigma)=1+\frac{k l_s^2}{|R^2+\sigma^2|}
\end{equation}
which can be seen to be nowhere singular. As a result, the constraint equation is modified slightly to become:
\begin{equation}
\frac{kl_s^2}{|R^2+\sigma^2|}(1-F^2) -F^2  \ge \frac{1}{\tilde{E}^2}-1.
\end{equation}
The constraint conditions, however, are essentially the same as those for the case of motion in the plane, the slight difference
being that the lead term never blows up. 


Now following \cite{kutasov} we define the effective potential in the general case to be
\begin{equation} \label{eq:veff}
\rm V_{eff} = \frac{-(1-F^2)}{H(\rho,\sigma)} + \frac{1}{H(\rho, \sigma)^2\tilde E^2}
\left( 1+ \frac{\tilde{L}_{\theta}^2}{\rho^2} + \frac{\tilde{L}_{\phi}^2}{\sigma^2} \right)
\end{equation}

At this point  we would like to investigate the equations of motion of the probe brane in this background, for simplicity
and convenience we consider each case seperately.

\subsection{Probe motion in the ring plane.}

As discussed previously, the ring plane is identified with the coordinates $\rho$ and $\theta$, and the harmonic
function is given by (\ref{eq:planeharmonic}). In the 'throat' geometry of the ring solution we can neglect the
factor of unity in $H(\rho)$ provided that $kl_s^2 >> \rho$.
Because of the distribution of the $NS5$-branes, we can consider probe motion either inside or outside of the ring.
Of course, the full equations of motion are complicated and need to be solved numerically (see later), but we can make
some progress by considering various limits. We first imagine $\rho << R$, which puts the probe brane very close to the
centre of the ring, and so we neglect the factor of $\rho$ in the expression for the harmonic function. The equation of
motion now reads
\begin{equation} \label{eq:throateom1}
\dot{\rho}^2=\frac{(1-F^2)R^2}{kl_s^2}-\frac{ R^4}{\tilde{E}^2 k^2l_s^4}.
\end{equation}
Which has a solution given by
\begin{equation} \label{eq:throatsoln1}
\rho = \frac{Rt}{\sqrt{k}l_s} \sqrt{(1-F^2)-\frac{ R^2}{\tilde{E}^2 k l_s^2}},
\end{equation}
i.e is linear in the bulk time $t$. Thus at $t$=0 we expect the probe to be at the centre of the ring, which is the furthest distance
from the $NS5$ branes, and as time evolves it moves outwards. Obviously (\ref{eq:throatsoln1}) will only be valid
in the small $\rho$ regime and so this solution cannot be trusted as the probe nears the ring. Furthermore, we can see that the solution becomes time independent if the following constraint is satisfied
\begin{equation}
(1-F^2) = \frac{R^2}{\tilde E^2 k l_s^2},
\end{equation}
and the probe will always remain at the origin. 

We can also consider
the regime where the probe is located far from the ring, i.e $\rho >> R$, but with $kl_s^2$ still larger than $\rho$.
The equations of motion are now modified slightly to become
\begin{equation} \label{eq:throateom2}
\dot{\rho}^2=\frac{(1-F^2)\rho^2}{kl_s^2}-\frac{\rho^4}{\tilde{E}^2 k^2l_s^4},
\end{equation}
which gives us the solution
\begin{equation}
\frac{1}{\rho}=\frac{1}{\tilde{E}l_s\sqrt{k(1-F^2)}}{\rm cosh }{\left(\frac{t\sqrt{1-F^2}}{\sqrt{k}l_s}\right)}
\end{equation}
This is the same expression that Kutasov found in \cite{kutasov} for a probe moving in the background of a stack of
coincident $NS5$-branes, and reinforces our claim that the ring distribution appears pointlike at large distances.
The above solution informs us that at $t=0$ the probe is at its maximum distance from the sources, and as time evolves it
is gravitationally attracted towards the ring. Of course, we must be aware that this solution is no longer valid in the
regime where the probe is near the ring.

So far we have made decent progress by simply considering the limits of the solutions, but in order to understand the ring
background we must try and find explicit solutions for the equation of motion in the region close to the ring. In order
to do this we have resorted to a numerical approach.

Consider first the case with  $L_\theta = L_\phi=0$. Figs 1 and 2 show numerical solutions for the distance
 $\rho(t) $. In Fig 1, we have taken the dimensionless energy density $\tilde{E} = 0.6 $ and the electric
flux $F=0$ or $F=0.8 $ and assumed a
positive initial probe brane velocity and a starting value of $\rho$ outside the ring.
(In this and all subsequent plots we have taken $ k l_s^2 / R^2 = 1$ for simplicity).
It is clear that in this case trajectories of the probe brane are bound to the ring and cannot escape to
infinity. The effect of turning on the electric flux on the probe is to increase its 'effective mass'
which results in the  maximum distance away from the ring being reduced.

\begin{figure}[htbp]
\vspace{0.5cm}
\centering
\epsfig{file=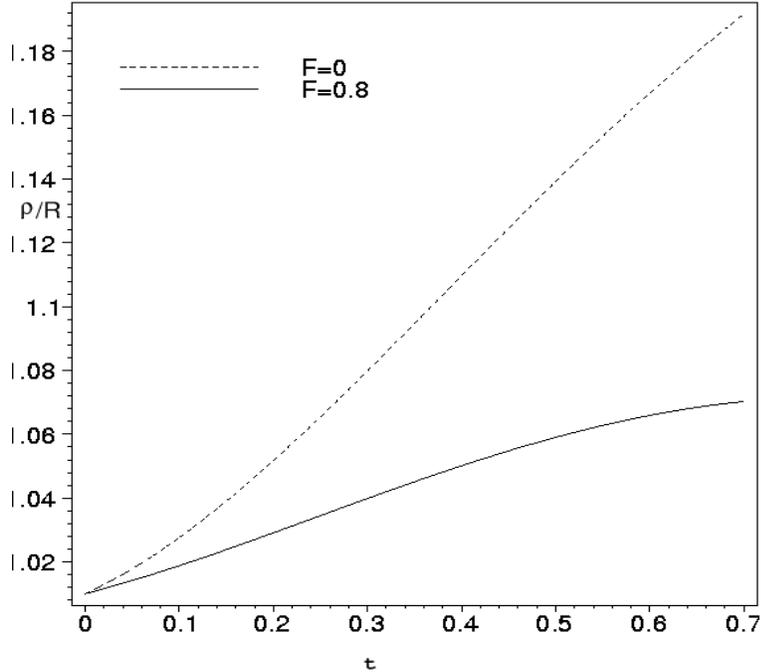, width= 10cm, height = 10cm}
\caption{Plot of the radial coordinate $\rho$ vs $t$, for $\tilde{E}=0.6, L_\theta= L_\phi =0$ and
taking electric flux $F =0$, $ 0.8$.} \label{fig: 1}
\end{figure}

 Fig 2 is the same situation but
with $\tilde{E} = 1.5 $. In this case both solutions describe a probe that can escape the ring and move
to infinity, the case with no electric flux having a greater escape velocity. Similar plots  for starting
positions inside the ring  ( or trajectories where the initial velocity
is towards  the ring starting from  $\rho > R $ ) show trajectories that eventually hit the ring
at $\rho = R$ although strictly speaking, one cannot follow these trajectories right to the ring location
as in this region there are large string effects.

\begin{figure}[htbp]
\vspace{-1cm}
\centering
\epsfig{file=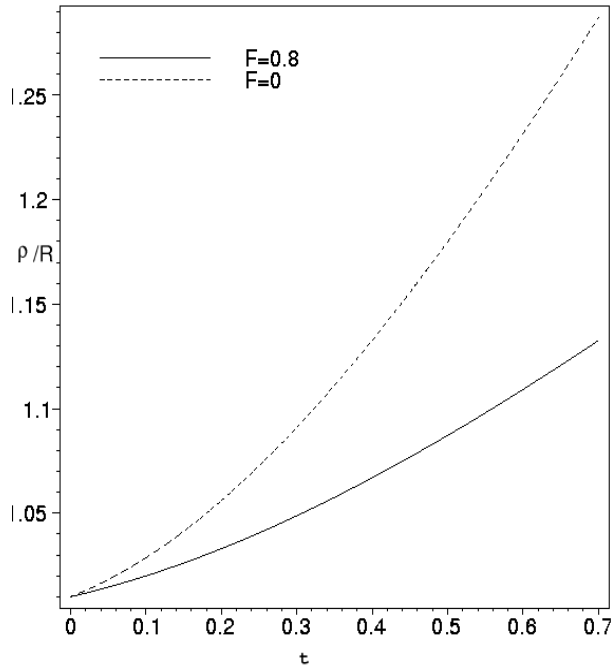, width= 8cm, height = 10cm}
\caption{Plot of the radial coordinate $\frac{\rho}{R}$ vs $t$, for $\tilde{E}=1.5, L_\theta= L_\phi =0$ and
taking electric flux $F =0$, $ 0.8$.} \label{fig: 2}
\end{figure}

These solutions can be understood in terms of the effective potential $V_{eff} (\frac{\rho}{R} ) $
plotted for various values of  $\tilde{E} $ and $F$. Fig 3 shows four such plots, taking
e.g. $\tilde{E} = 0.6 $ or $1.5$ and $ F=0 $ or $0.8$. These plots cover the region from $\rho=0 $
at the centre of the ring, to values outside.

\begin{figure}[t]
\vspace{-6.5cm}
\centering
\epsfig{file=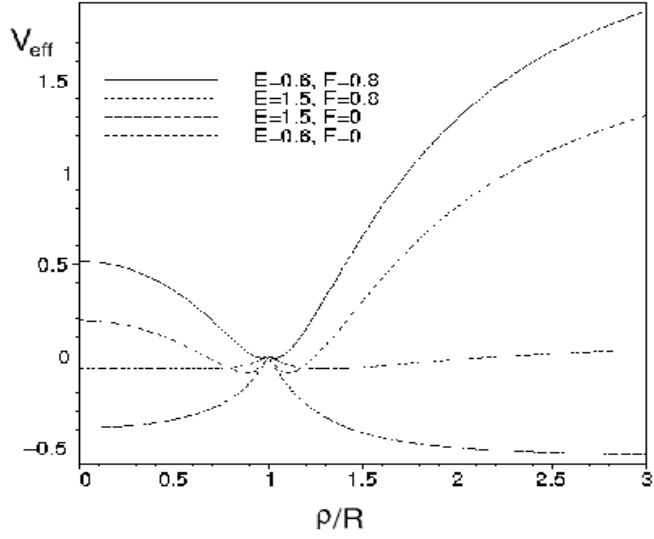, width= 12cm, height = 15cm}
\caption{Plot of the effective potential $V_{eff}$ vs $ \frac{\rho}{R}$, for $\tilde{E}=(0,1.5), L_\theta =
L_\phi =0$ and taking electric flux $F = 0.6$, $ 0.8$. } \label{fig: 3}
\end{figure}

\subsection{Probe motion transverse to the ring plane.}
As in the previous section we initially consider the situation when $\sigma << R$. The equation of motion in this plane
is similar to the one for motion in the ring plane, and we obtain the solution
\begin{equation}\label{throateom3}
\sigma = \frac{Rt}{\sqrt{k}l_s}\sqrt{(1-F^2)-\frac{R^2}{\tilde{E}^2 k l_s^2}}.
\end{equation}
The same comments apply to this solution, except that in this instance the probe brane is no longer moving towards the
$NS5$ branes as time evolves, since it is moving in the transverse plane to the ring. Again we must be aware that this
linear solution is only valid for small $\sigma$.
If we now consider the case where $\sigma >> R$, then we can again imagine that at large enough distances the ring
distibution will appear pointlike and we expect to recover a similar solution to the previous section. This is indeed the case, and
the solution is
\begin{equation}\label{eq:throateom4}
\frac{1}{\sigma}=\frac{1}{\tilde{E}l_s\sqrt{k(1-F^2)}}{\rm cosh }{\left(\frac{t\sqrt{1-F^2}}{\sqrt{k}l_s}\right)}
\end{equation}
Where the same comments must apply when considering the critical value of the electric field.

Once again we can understand the solutions in between small or large values of $\sigma /R $ by
using numerical methods. Given that a probe is
attracted to the NS5 ring if it is postioned above it,  we might guess that a brane, with small enough energy,
falling towards the centre of the ring from above the plane of the ring, would pass through its centre and then extend
below it to some maximum distance and then be attracted back through the centre of the ring and so on. That is we might expect
a special solution describing the oscillatory motion of the probe through the ring centre. Such a solution should match
 on to the linear solution described above when the probe is at a small distance either above or below the ring plane,
ie when $\sigma /R  << 1 $.

Fig4 shows a plot of the numerical solution in this case
\begin{figure}[t]
\vspace{-2.5cm}
\centering
\epsfig{file=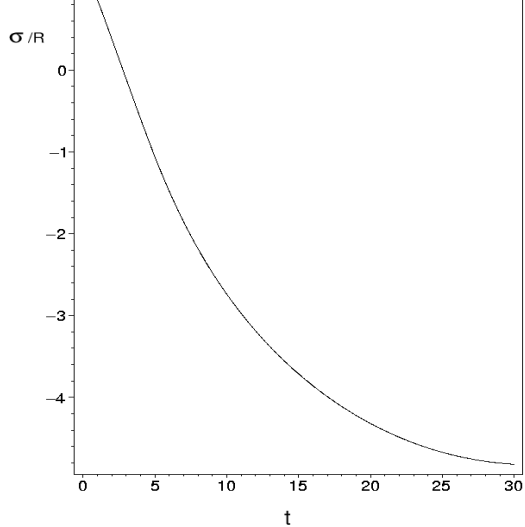, width= 8cm, height = 8cm}
\caption{Plot of the distance $\frac{\sigma}{R}$ of the probe from the ring plane vs $ t $, for $\tilde{E}=0.98 , L_\theta =
L_\phi =0$ and taking electric flux $F =0$. This describes motion of the probe through the centre of the ring at
$\rho = 0 $  } \label{fig: 4}
\end{figure}

The linear behaviour of $\sigma (t) $ as a function of $t$, for small values of $\sigma /R$ is evident from Fig 4 so matches our analytic solution above.
In this plot the  probe starts from $\frac{\sigma}{R} = 1.3 $ at $t=0$ and reaches a maximum distance below the ring of 
about $4.5 R$.
After this time the probe is attracted back up through the ring and the process repeats. The motion thus describes oscillation between the two zeros
of the effective potential $ V_{eff} ( \sigma ) $ which is plotted in Fig 5.
What is also noteworthy about this particular solution is that it is stable to stringy corrections since we can control 
the minimum distance the probe comes to the  ring by making the ring radius sufficiently large.

\begin{figure}[t]
\vspace{-2.5cm}
\centering
\epsfig{file=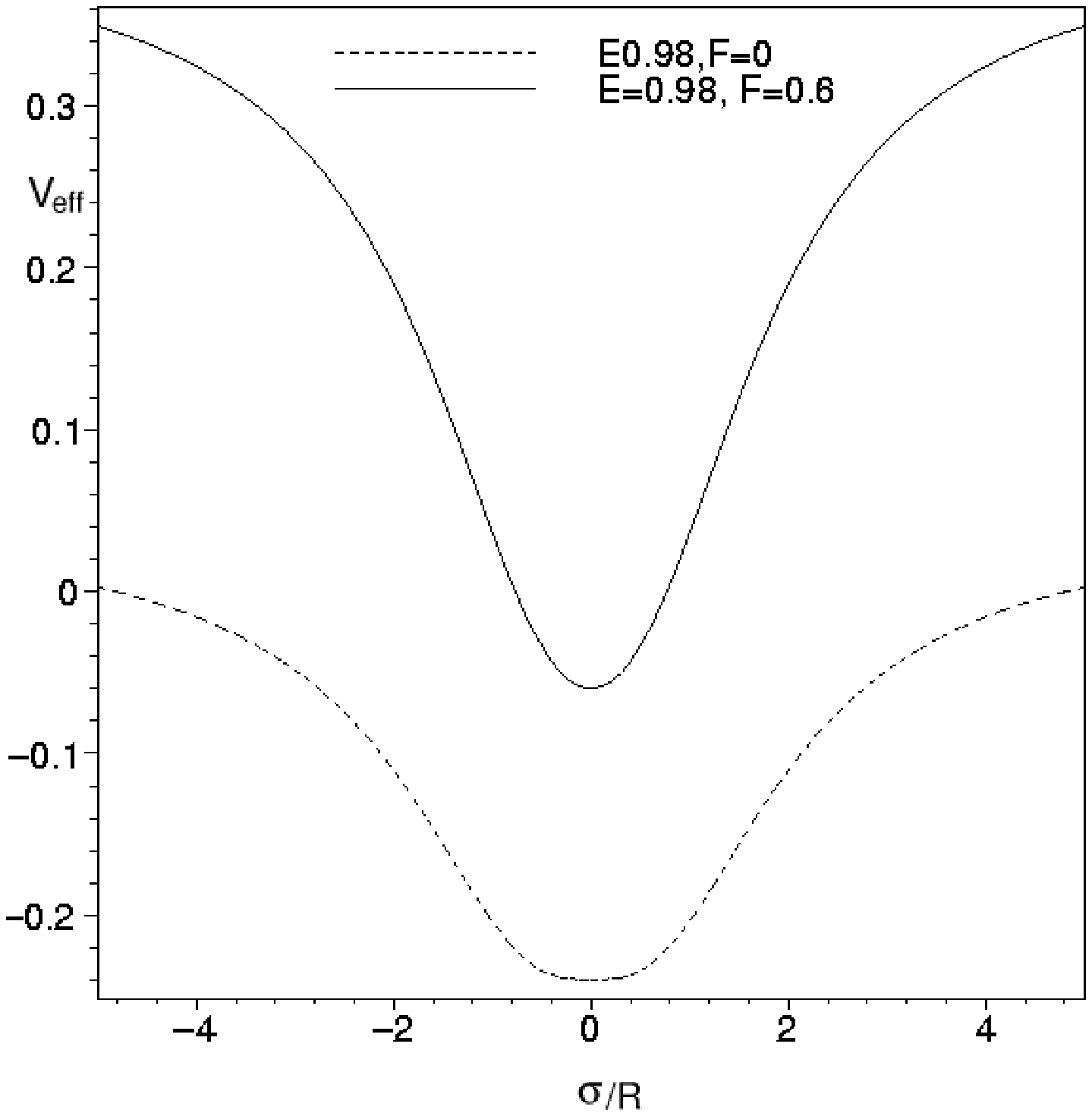, width= 8cm, height = 10cm}
\caption{Plot of $V_{eff} (\rho = 0, \sigma ) $ relevant to the study of probe motion through the centre of the ring.
$\tilde{E}= 0.98, L_\theta =
L_\phi =0$, whilst the electric flux is taken to be $F =0, 0.6$. } \label{fig: 5}
\end{figure}

In the plot of Fig 5 we have also shown the effect turning on some electric flux has on $V_{eff} $. Its clear that it results in making this
potential more postive everywhere and hence reduces the range and period of the oscillation through the center. What is particulary interesting is
that there exists a critical value of the flux $F$ (around $0.6$ ) beyond which oscillation is not possible and the probe is stuck at the ring center.
The existence of this critical value of the flux $F$ (for a given energy density $\tilde{E} $ ) can easily be understood.

The energy density  of a probe brane  carrying flux $F$ which is at rest above the ring plane at distance $\sigma $
(with $\rho = 0 $) must satisfy
\begin{equation}
\frac{1}{\tilde{E}^2 } - (1-F^2 ) H(R,\sigma ) = 0
\end{equation}

where the harmonic function $H(R,\sigma ) $ is given in (\ref{eq:perpharmonic}). Now we see that for given
$\tilde{E} $, turning on the flux, $F$, on a probe brane which was initially at some point $\sigma $ above the ring plane
means that in order to satisfy this equation the probe has to move closer to the plane, thus increasing the value of
 $H(R, \sigma )$. But as we keep increasing $F$ this cannot carry on indefinitely as there is a maximum value
 that $H(R, \sigma ) $ can take ( for fixed value of $ k_s l_s^2 /R^2 $ ), which is its value at the centre of the ring.
 Thus there is a critical value of flux for a given $ \tilde E $. Of course one has to bear in mind that we cannot make
 the factor of $1-F^2 $ too small as our derivations are only perturabtive in $F$.

So far we have only considered radial trajectories with vanishing angular momentum, at this point we must also consider
the probe dynamics when the momenta are non zero.

\subsection{Motion in the ring plane with $\tilde{L}_{\theta} \neq 0$ }

If we retain the angular momentum term in (\ref{eq:eom}) we must try to solve
\begin{equation}
\dot{\rho}^2 = \frac{(1-F^2)}{H(\rho )}-\frac{1}{\tilde{E}^2 H(\rho)^2} \left( 1 + \frac{\tilde{L}_{\theta}^2 }{\rho^2} \right)
\end{equation}
As in the previous sections we begin by considering the limit $\rho << R$, which puts the probe brane inside the ring. 
We find that the solution in this instance is given by the following expression;
\begin{equation}
\rho^2 = R^2\frac{t^2(\tilde{E}^2 k l_s^2(1-F^2))^2-2 R^2 \tilde{E}^2kl_s^2t^2(1-F^2)+R^4t^2 +
 \tilde{E}^2k^2l_s^4 \tilde{L}_{\theta}^2}{\tilde{E}^2k^2l_s^4(\tilde{E}^2kl_s^2(1-F^2)-R^2)} 
\end{equation}
This somewhat complicated expression reduces to (\ref{eq:throatsoln1}) in the limit $\tilde{L}_{\theta}=0$.
At the opposite end of the spectrum in the $\rho >> R$ regime, we find that the solution is given by
\begin{equation}
\frac{1}{\rho} = \frac{1}{\tilde{E}\sqrt{k}l_s\sqrt{(1-F^2)-\tilde{L}_
{\theta}^2/kl_s^2\tilde{E}^2 }}{\rm  cosh} \left( \frac{t}{\sqrt{k}l_s}\sqrt{(1-F^2)-
\frac{\tilde{L}_{\theta}^2}{kl_s^2\tilde{E}^2 }} \right) 
\end{equation}
which again reproduces the earlier result in the limit of no angular momentum, and shows us that
the momentum term has the effect of slowing the decrease of $\rho$ in the $t \to \infty$ limit.
Furthermore this equation provdes us with bounds on the angular momentum, since it must
satisfy the constraint
\begin{equation}
\tilde L_{\theta}^2 < (1-F^2) k l_s^2 {\tilde E}^2.
\end{equation}
If $\tilde L_{\theta}$ saturates this bound then the only solution is $1/\rho = 0$. Thus we see
that increasing the flux automatically leads to a reduction in the angular momentum. This is in
agreement with our intuative picture of the flux providing extra mass on the brane.

In order to study trajectories of the full theory without resorting to the special limits in $\rho $
discussed above we again look to numerical solutions. We expect that solutions to the full theory will
describe the probe brane
in an unstable orbit about the ring. This is confirmed in Fig 6 which is a  parameteric plot in the
$(\rho ,\theta )$ plane of a solution which starts at ($\frac{\rho}{R} =1.1 , \theta = 0 $) at $t=0$.
In this plot we took $\tilde{E} = 1.02, \frac{L_\theta}{R}=0.98 $ and $F=0.2$ and we see the trajectory 
spiralling outwards
from the ring.
Starting with different initial conditions would produce e.g. trajectories that spiral towards the ring
(either starting from inside or outside) and eventually end on there.
\subsection{Motion transverse to the ring plane with $\tilde L_{\phi} \ne 0$.}
\begin{figure}[htbp]
\vspace{-0.5cm}
\centering
\epsfig{file=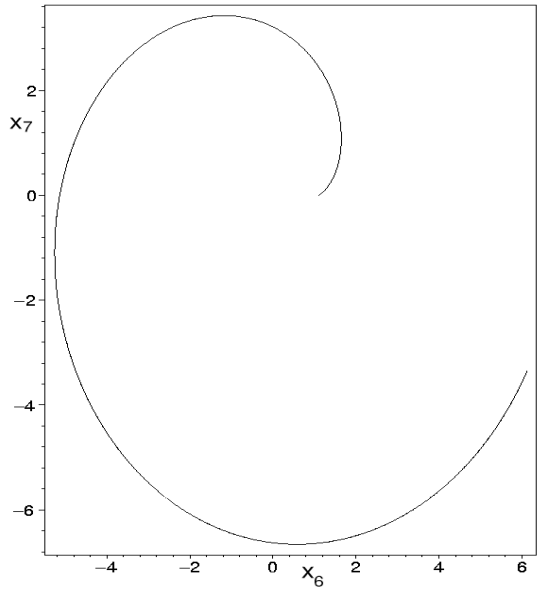, width= 8cm, height = 8cm}
\caption{Plot of brane trajectory in the $ x_6 - x_7 $ (ie $\rho, \theta $ )  plane, for $\tilde{E}=1.02$,
$\frac{L_\theta}{R}=0.98 $ and taking electric flux $F =0.2$ } \label{fig: 6}
\end{figure}

Fig 7 shows the plot of $V_{eff} $ vs $\frac{\rho}{R} $ with all other values as above.  The same function is
also shown for $\tilde{E} = 0.75 $, in which case the spiral trajectories cannot escape to infinity.

\begin{figure}[htbp]
\vspace{-3cm}
\centering
\epsfig{file=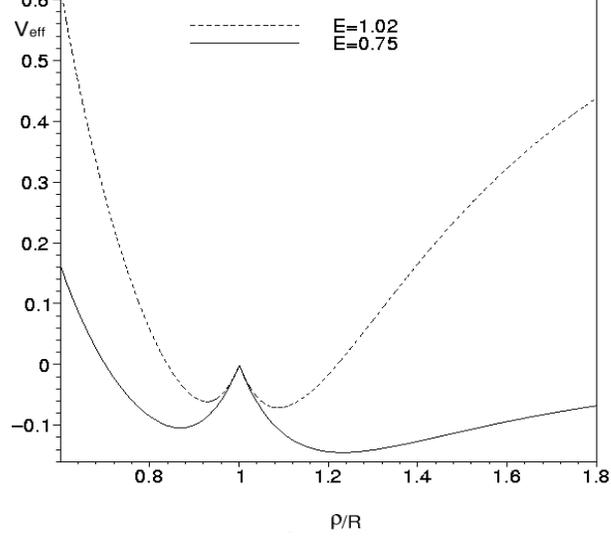, width= 8cm, height = 8cm}
\caption{Plot of  $V_{eff} vs \frac{\rho}{R} $ with $\tilde{E}= (0.75, 1.02)$,
$\frac{L_\theta}{R}= 0.98 $ and taking electric flux $F =0.2$ } \label{fig: 7}
\end{figure}

\begin{figure}[htbp]
\vspace{-2.5cm}
\centering
\epsfig{file=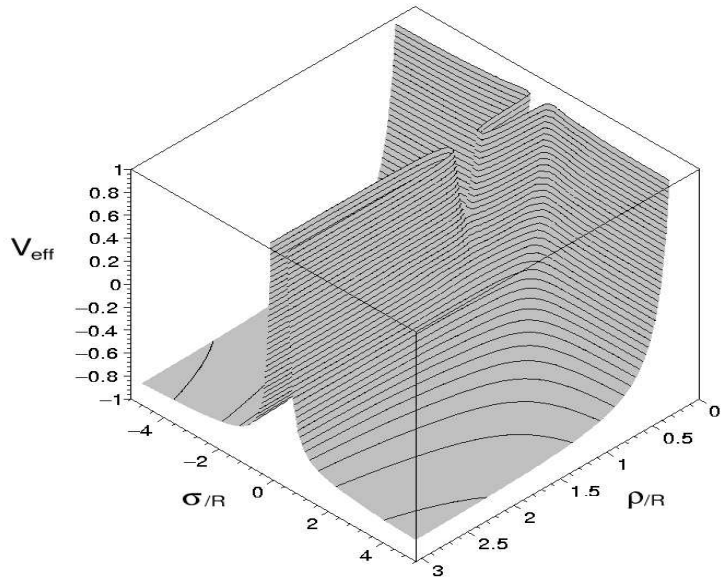, width= 12cm, height = 12cm}
\caption{A 3-d plot of  $V_{eff}$ vs $\frac{\rho }{R} $ and $\frac{\sigma}{R}$ with $\tilde{E}= 3.29$,
$\frac{L_\theta}{R}= 1.2 $, $\frac{L_\phi}{R}=1.4 $ and taking electric flux $F =0$ } \label{fig: 8}
\end{figure}

Figure 8 shows a 3-d plot of the effective potential for non-zero values for $L_\theta $ and $L_\phi $
and  $\tilde{E} = 3.29 $ respectively. For this value of the energy we expect trajectories corresponding
to the probe brane moving away from the ring to escape to infinity, which corresponds to  $V_{eff}$ becoming negative
at large distances in $\sigma , \rho $ as can be seen in the plot. On the other hand probe branes moving towards the ring
feel a generic repulsion due to the presence of a centrifugal barrier coming from the angular momentum terms in
$V_{eff} $.  This would lead to scattering of the probe brane off the ring which happens also in the case of point like 
source of NS5-branes, in the presence of angular momentum  \cite{kutasov}. The plot also shows a `gap' in the 
centrifugal barrier located at the ring location $\sigma = 0, \rho = 1 $ so that its possible for some trajectories to 
still end on the ring itself (ignoring possible stringy corrections). A numerical study is needed to distinguish these 
various possibilities. Unfortunately this requires solving the full set of non-linear equations of motion for 
$\rho, \sigma, \theta $ and $\phi $ which requires methods that go beyond those we used earlier.   
Nevertheless it would be interesting to explore the nature of trajectories in this case.

\vspace{2cm}
\section{Discussion}
In this paper we have discussed same basic properties of probe-brane dynamics in the background geometry of 
a continuous distribution of NS5 branes around a ring. We have found  similarities to the situation when the NS5 branes are 
concentrated at a point, namely that there is generally an attaction of the probe resulting in some trajectories 
ending on the ring or else describing the scattering of the probe off the ring. We have found an interesting solution
where the probe can oscillate through the centre of the ring in a manner that is stable to stringy corrections 
if the ring is taken large enough. Our studies involved a mixture of analytical and numerical methods.

A number of future problems concerning the properties of D-branes moving in the NS5-ring geometry come to mind. 
The fate of probe branes falling towards the ring could be studied using similar methods to those developed 
in \cite{decay}, \cite{NST} and \cite{sahayakan}. Since the 
NS5-brane ring geometry is rather different from that due to a point source we might expect different behaviour as regards the probes 
decay into closed string modes.  Indeed the decay crossection calculation would doubtless make use of the 
exact description of the system in terms of the $SU(2)/U(1) \times SL(2,\mathbf{R})/U(1) $ CFT . In this context
it would be interesting to study the stability of the transverse oscillating solution discussed earlier, 
(when the probe moves through the ring centre) to decay into closed string modes. 
(\cite{sfetsos},\cite{sfetsos2}).

\end{document}